\documentclass[twocolumn]{aastex701}
\usepackage{amsmath}

\begin{document}

\title{Impact of Nuclear Level Density on $\mathit{r}$-process Rare-Earth Peak Nucleosynthesis}

\author{Hang Xu, Peng-Xiang Du}
\email{}
\affiliation{%
	College of Physics, Jilin University, Changchun 130012, China
}%

\author{Jian Li}
\email[show]{jianli@jlu.edu.cn}
\affiliation{%
College of Physics, Jilin University, Changchun 130012, China
}%
\author{Dong-Liang Fang}
\email[show]{dlfang@impcas.ac.cn}
\affiliation{%
    Institute of Modern Physics, Chinese Academy of Sciences, Lanzhou 730000, China
}%

\begin{abstract}

The rare-earth peak ($A\sim164$) is a prominent feature of the $\mathit{r}$-process, and previous theoretical studies suggest that it is possibly linked to local nuclear structural effects. However, the nuclear level density (NLD)—a physical quantity directly reflecting these properties—has been largely overlooked compared to other structural properties such as nuclear masses. To address this, we perform $\mathit{r}$-process simulations across three astrophysical scenarios using neutron-capture rates derived from six distinct NLD models. Our results reveal that microscopic models yield systematic deviations in NLD relative to phenomenological ones, leading to critical impacts on nucleosynthesis. Specifically, systematic NLD differences in even-$A$ nuclei redirect the nuclear flow, accelerating the early formation of the rare-earth peak and temporarily enhancing its magnitude. This underlying structural shift also fundamentally alters the $\mathit{r}$-process sensitivity to the neutron-capture rate, effectively eliminating its dependence on the odd-even nature of protons. Overall, these findings demonstrate that the internal nuclear structure encoded within NLDs can collectively induce a global redirection of the nucleosynthesis pathway, highlighting the critical need for self-consistent microscopic inputs in future simulations.

\end{abstract}


\section{Introduction}

The rapid neutron-capture process ($\mathit{r}$-process) is responsible for the synthesis of approximately half of the elements heavier than iron in the universe. First proposed in the seminal work by \citet{burbidgeSynthesisElementsStars1957}, the $\mathit{r}$-process occurs under conditions of extreme neutron densities ($n_n \gtrsim 10^{20} \text{ cm}^{-3}$) and high temperatures ($T \sim 10 \text{ GK}$), where rapid neutron captures compete with $\beta$-decays and photodissociation to drive the nuclear flow toward the extremely neutron-rich isotopes.

Identifying the astrophysical sites for the $\mathit{r}$-process remains a major challenge. While neutrino-driven winds from core-collapse supernovae (CCSNe) were initially favored, subsequent simulations indicated that they generally lack the necessary physical conditions (e.g., sufficiently low electron fraction) to synthesize heavy $\mathit{r}$-elements \citep{woosleyRProcessNeutrinoheatedSupernova1994, qianNucleosynthesisNeutrinoDrivenWinds1996}. Current theories therefore propose that rare, extreme scenarios, such as magnetorotational supernovae and collapsars, may host the necessary conditions for heavy $\mathit{r}$-process nucleosynthesis \citep{wintelerMagnetorotationallyDrivenSupernovae2012,mostaRprocessNucleosynthesisThreedimensional2018,siegelCollapsarsMajorSource2019}. Meanwhile, binary neutron star mergers (BNSMs) emerged as robust production sites due to their naturally neutron-rich ejecta \citep{hotokezakaNeutronStarMergers2018,gorielyRPROCESSNUCLEOSYNTHESISDYNAMICALLY2011}. This was confirmed by the multi-messenger observation of GW170817 \citep{abbottMultimessengerObservationsBinary2017}, which provided the first direct evidence of $\mathit{r}$-process synthesis in these environments \citep{chenNeutronStarMergers2024}.

Another major challenge in $\mathit{r}$-process studies arises from the uncertainties in nuclear physics inputs. Owing to the lack of measurements of reaction rates for extremely neutron-rich nuclides, simulations rely heavily on theoretical extrapolations of these rates. These extrapolations exhibit increasing divergence toward the neutron drip line, where model predictions can differ by several orders of magnitude. Extensive sensitivity studies have quantified the impact of these uncertainties, covering nuclear masses \citep{mumpowerImpactIndividualNuclear2015,martinImpactNuclearMass2016,maInfluenceNuclearMass2019}, neutron-capture rates \citep{surmanSensitivityStudiesWeak2014}, $\beta$-decay rates \citep{mumpowerSensitivityStudiesMain2014}, and fission yields \citep{martinez-pinedoRoleFissionRprocess2007}. While current models can generally reproduce the solar $\mathit{r}$-process abundance distribution, achieving precise descriptions requires more robust nuclear data \citep{mumpowerImpactIndividualNuclear2015}, making the alignment between calculated and observed abundances a critical test for validating theoretical models. The two most prominent features of the solar $\mathit{r}$-process abundance pattern are the major peaks resulting from material accumulation around $A \sim 130$ and $A \sim 195$. These features originate from the nuclear shell structure associated with the neutron magic numbers $N=82$ and $N=126$, respectively \citep{burbidgeSynthesisElementsStars1957}.

In the intermediate mass region between these major peaks, a distinct substructure known as the rare-earth peak emerges. Given its specific location, the rare-earth peak must originate from a mechanism different from the standard magic numbers. It is generally suggested that the emergence of this feature may be related to nuclear deformation, sub-shell closures, or the deposition of fission fragments. However, the exact physical origin of the rare-earth peak remains an open question. Proposed formation mechanisms of the rare-earth peak include kinks in neutron separation energy ($S_n$) during a hot $\mathit{r}$-process \citep{engelSourceRareEarthElement1997}, dynamic formation scenarios in a cold $\mathit{r}$-process \citep{mumpowerFormationRareearthPeak2012}, and late-time fission deposition \citep{gorielyFundamentalRoleFission2015}. More recently, \citet{vasshNeedLocalNuclear2022} showed that local nuclear structure effects in the lanthanide region play a dominant role in shaping the rare-earth peak. Complementing this, systematic sensitivity studies have further investigated the specific impacts of fission, nuclear masses, and $\beta$-decay rates on the rare-earth peak formation \citep{haoInfluenceSpontaneousFission2022,haoSensitivityRprocessRareearth2023,haoImpactNuclear$ensuremathbeta$decay2023}. Meanwhile, the rare-earth peak serves as a unique probe for internal nuclear properties \citep{mumpowerRAREEARTHPEAK2012}. This sensitivity has enabled reverse-engineering approaches to predict nuclear masses, predictions that were subsequently validated by precision measurements \citep{mumpowerLINKRAREEARTHPEAK2016,orfordPrecisionMassMeasurements2018}.

NLD serves as a fundamental input parameter for calculating nuclear reaction cross sections via the Hauser-Feshbach formalism \citep{hauserInelasticScatteringNeutrons1952,gorielyImprovedPredictionsNuclear2008}. Based on the compound nucleus model and Fermi's Golden Rule, the NLD strongly influences the probability of the compound nucleus de-exciting via $\gamma$-decay. Meanwhile, under specific astrophysical scenarios, the formation of the rare-earth peak is believed to be related to the evolution of neutron-capture rates \citep{mumpowerInfluenceNeutronCapture2012,mumpowerFormationRareearthPeak2012}, which depend on various nuclear properties. While the impact of nuclear masses has been extensively studied, the NLD—which encodes vital structural information such as sub-shell closures and shape evolution—has frequently been oversimplified in previous research. This is particularly significant given that the formation of the rare-earth peak likely has a strong connection to these nuclear structures.

NLD models are generally categorized into phenomenological and microscopic approaches. In this work, we use six representative NLD models in the analysis. These models provide open access datasets covering the thousands of neutron-rich nuclei required for $\mathit{r}$-process calculations. For the phenomenological approaches, we employ the Back-Shifted Fermi Gas Model (BFM) \citep{dilgLevelDensityParameters1973}, the Constant Temperature Model (CTM) \citep{gilbertCompositeNuclearlevelDensity1965}, and the Generalized Superfluid Model (GSM) \citep{koningGlobalLocalLevel2008}. These models are based on the Fermi gas approximation with parameters calibrated to experimental data. For the microscopic approaches, we employ the Hartree-Fock Bardeen-Cooper-Schrieffer statistical model (HFBCS+stat.) \citep{demetriouMicroscopicNuclearLevel2001}, the Hartree-Fock-Bogoliubov combinatorial model (HFB+comb.) \citep{gorielyImprovedMicroscopicNuclear2008}, and the temperature-dependent Gogny-Hartree-Fock-Bogoliubov combinatorial model (THFB+comb.) \citep{hilaireTemperaturedependentCombinatorialLevel2012}. These microscopic approaches have the major advantage of being able to treat shell, pairing, and deformation effects in a consistent way. This consistency is particularly crucial for capturing the delicate structural details required to investigate the formation mechanism of the structure-sensitive rare-earth peak.

Prior investigations into nuclear physics inputs have predominantly relied on phenomenological NLD models, such as the BFM. These phenomenological models allow NLDs to be rapidly calculated via simple analytic formulae when different nuclear masses are adopted. However, this mathematical convenience comes with significant physical limitations. In particular, for models such as the BFM, a highly simplified energy dependence for the key NLD parameter $a$ is employed, which leads to an inaccurate accounting of nuclear spectra. Besides, the complicated pairing effects are simplified as energy-shift terms \citep{arnouldRprocessStellarNucleosynthesis2007}. With the inclusion of limited numbers of free parameters, these models can successfully reproduce available experimental data. However, such extrapolations become unreliable when describing the excitation spectra of neutron-rich nuclei required for the $\mathit{r}$-process, where experimental data are absent. \citet{mumpowerInfluenceNeutronCapture2012} investigated the impact of neutron-capture rates on the rare-earth region, focusing primarily on individual nuclear properties and environmental effects. Nevertheless, they paid less attention to the underlying nuclear physics inputs. While recent studies have noted the potential influence of NLDs on the rare-earth peak \citep{poglianoImpactLevelDensities2023}, a systematic understanding of the physical mechanisms is still lacking. Therefore, the primary goal of this work is to investigate how the specific nuclear structure information embedded within NLDs affects the formation of the rare-earth peak. We systematically examine the impact of neutron-capture rates, calculated using six distinct NLD models, on the peak formation across three representative astrophysical scenarios. Furthermore, we perform an NLD-based neutron-capture rate sensitivity study and investigate how the fundamental differences between microscopic and phenomenological models reshape the sensitivity landscape of the rare-earth peak.

The paper is organized as follows. Section \ref{sec:methods} details the astrophysical environments, nuclear physics inputs, and the methodology of the sensitivity analysis employed in this study. In Section \ref{sec:results}, we discuss the impact of different NLD models on the formation of the rare-earth peak and analyze how the variation of NLD models influences the sensitivity of the rare-earth peak to neutron-capture rates. Finally, Section \ref{sec:summary} provides a summary of our results and an outlook for future work.

\section{$\mathit{r}$-process Calculations} \label{sec:methods}
\subsection{$\mathit{r}$-Process model} \label{subsec:r_model}

In this work, we performed $\mathit{r}$-process network calculations using the SkyNet code \citep{Lippuner_2017}. Adapting the parameterized trajectories for neutron star merger (NSM) dynamical ejecta utilized by \citet{vasshMarkovChainMonte2021}, we simulated the abundance evolution of over 7000 nuclides from Nuclear Statistical Equilibrium (NSE) up to $10^9$~s. The calculations were initialized by defining the entropy per baryon ($s$), electron fraction ($Y_e$), dynamical timescale ($\tau$), and the temporal evolution of density. We explored three distinct outflow conditions: (1) hot ($Y_e = 0.18, s = 30~k_B/\text{baryon}, \tau = 70$~ms); (2) hot/cold ($Y_e = 0.2, s = 20~k_B/\text{baryon}, \tau = 10$~ms); and (3) cold ($Y_e = 0.2, s = 10~k_B/\text{baryon}, \tau = 3$~ms).

Regarding the electron fraction, we found that a value of $Y_e = 0.2$ failed to reproduce the third $\mathit{r}$-process peak in the hot scenario. Consequently, we adjusted $Y_e$ to 0.18 for this scenario. This adjustment ensures that the nuclear flow extends sufficiently into the trans-lead region, maintaining consistency with the other scenarios, while simultaneously preventing the onset of excessive fission cycling. The density evolution follows the default SkyNet trajectory profile:
\begin{equation}
\rho(t) = 
\begin{cases} 
\rho_0 \mathrm{e}^{-t/\tau}, & t < 3\tau  \\
\rho_0 \left(\frac{3\tau}{\mathrm{e}t}\right)^3, & t \ge 3\tau,
\end{cases}
\label{eq:density_profile}
\end{equation}
where $\rho_0$ is the initial density of the trajectory.

The $\mathit{r}$-process path is defined by the isotope with the maximum abundance for each element, serving as a robust tool to track the nuclear flow. As the system evolves, this path is driven toward the valley of stability by $\beta$-decays. Under corresponding environmental conditions, the path is dominated by specific nuclear physics inputs, such as nuclear masses or neutron-capture rates. When these dominant inputs exhibit non-linear variations along the $\mathit{r}$-process path, material accumulation occurs, potentially leading to the formation of abundance peaks such as the rare-earth peak. In the hot scenario, a high dynamical timescale maintains high temperatures and densities for an extended duration. Under these conditions, according to the principle of detailed balance, the $(n,\gamma) \rightleftharpoons (\gamma, n)$ equilibrium persists until freeze-out (defined as the point where the abundance of seed nuclei equals the neutron abundance). In this regime, nuclear masses directly determine the path via the Saha equation:\begin{equation}\frac{Y(Z, N+1)}{Y(Z, N)} \propto \frac{G(Z, N+1)}{2 G(Z, N)} \frac{N_n}{(k T)^{3/2}} \exp \left[\frac{S_n(Z, N+1)}{k T}\right],\label{eq:Saha_equilibrium}\end{equation}
where $G(Z,N)$ are the partition functions, $N_n$ is the neutron number density, $kT$ is the temperature in MeV, $S_n$ represents the difference in binding energy between the nuclei $(Z, N+1)$ and $(Z, N)$. Consequently, the rare-earth peak formation is primarily dominated by nuclear masses, with neutron-capture rates gaining influence only during the freeze-out phase when equilibrium breaks down. Conversely, in the cold scenario, the rapid temperature drop breaks equilibrium early, and the path is driven by the competition between neutron capture and $\beta$-decay, making the rare-earth peak highly sensitive to capture rates. The hot/cold scenario represents an intermediate regime where equilibrium is maintained longer than in the cold scenario but breaks earlier than in the hot scenario, resulting in a hybrid formation mechanism.

\subsection{Nuclear physics inputs} \label{subsec:inputs}

We employed the nuclear reaction code TALYS 1.96 \citep{koningTALYSModelingNuclear2023} to calculate neutron-capture rates using six distinct NLD models. Since the present work focuses on local nuclear structure effects in the rare-earth region, we chose $45\leq Z\leq87$ as the proton-number range for our calculations. This range was selected to include the proton magic numbers Z=50 and Z=82, thereby reducing artificial boundary effects near magic nuclei, while excluding the actinide region to limit the influence of fission-fragment deposition on the rare-earth peak.

In the statistical Hauser-Feshbach formalism, the stellar neutron-capture rate is obtained by averaging the reaction cross section over a Maxwell-Boltzmann distribution. It can be expressed as proportional to the integration of the cross section:
\begin{equation}
\begin{split}
& N_A \langle \sigma v \rangle (T) \propto \\
& \frac{1}{(k_B T)^{3/2}} \int_{0}^{\infty} \sum_{\mu} \frac{2 J_t^{\mu} + 1}{2 J_t^0 + 1} \sigma_{n\gamma}^{\mu}(E) E \exp\left [ - \frac{E + E_{\mu}}{k_B T} \right ]  dE,
\end{split}
\label{eq:reaction_rate_prop}
\end{equation}
where $T$ is the temperature, $k_B$ is the Boltzmann constant, $J_t$ is the target's spin, $E$ is the relative energy between the target and the neutron, and $\mu$ and 0 represent the $\mu$-th excited state and the ground state, respectively. The total neutron-capture cross section, $\sigma_{n,\gamma}$, is the sum over all possible compound nucleus (CN) spin ($J$) and parity ($\pi$) states:
\begin{equation}
\sigma_{n,\gamma}(E_{x}) = \sum_{J,\pi} \sigma^{\text{CN}}(E_{x}, J, \pi) \frac{\mathcal{T}_\gamma(E_{x}, J, \pi)}{\mathcal{T}_{\text{tot}}(E_{x}, J, \pi)}.
\label{eq:cross_section_mechanism}
\end{equation}
Here, $\sigma^{\text{CN}}$ represents the compound nucleus formation cross section and is determined by the optical potential. The second factor represents the decay probability, determined by the ratio of the $\gamma$-transmission coefficient ($\mathcal{T}_\gamma$) to the total transmission coefficient ($\mathcal{T}_{\text{tot}}$). Crucially, the NLD ($\rho_{nld}$) enters directly into the calculation of $\mathcal{T}_\gamma$, serving as the weighting function for the available decay channels:
\begin{equation}
\begin{split}
&\mathcal{T}_\gamma(E_{x}, J, \pi) = \\
&2\pi \sum_{X,L} \int_0^{E_{x}} E_\gamma^{2L+1} f^{XL}(E_\gamma) \rho_{\mathrm{nld}}(E_{x} - E_\gamma, J, \pi) dE_\gamma,
\end{split}
\label{eq:transmission_coeff}
\end{equation}
where $f^{XL}$ is the $\gamma$-ray strength function with multipolarity $XL$, and $E_{\gamma}$ is the $\gamma$-ray energy. Since the kinetic energy of neutrons in $\mathit{r}$-process environments is negligible compared to $S_n$, the compound nucleus is formed at an excitation energy $E_x \approx S_n$. Combining Eqs.~(\ref{eq:reaction_rate_prop})--(\ref{eq:transmission_coeff}), the integrated NLD below $S_n$ serves as a key weighting factor for the final neutron-capture rate.

For the NLD inputs, we investigated two classes of models available in TALYS, namely the phenomenological ones and the microscopic ones. The former (CTM, BFM and GSM) are rooted in the Fermi gas framework. They differ primarily in their functional descriptions of the level density’s energy dependence: the CTM matches a constant temperature formula at low energies to a Fermi gas model at higher energies; the BFM employs the back-shifted Bethe formula across the entire energy range; the GSM incorporates collective enhancements and accounts for the phase transition from the superfluid state to the gas phase. In contrast, the latter (HFBCS+stat., HFB+comb. and THFB+comb.) derive single-particle levels from effective nucleon-nucleon interactions, such as Skyrme or Gogny forces, explicitly accounting for the nuclear deformation. In these frameworks, pairing correlations are treated using Bardeen-Cooper-Schrieffer (BCS) or Bogoliubov formalisms, and the final level densities are computed using either statistical \citep{demetriouMicroscopicNuclearLevel2001,bezbakhLevelDensitiesHeaviest2014} or combinatorial \citep{gengCalculationMicroscopicNuclear2023,hilaireGlobalMicroscopicNuclear2006,uhrenholtCombinatorialNuclearLeveldensity2013} methods. To ensure consistency in the binding energies and separation energies used across all calculations, the Finite Range Droplet Model (FRDM2012) \citep{mollerNuclearGroundstateMasses2016} was employed as the underlying mass model for all six NLD cases. For the $\gamma$-ray strength function, although NLDs and GSFs originate from the same underlying nuclear structure physics, our additional tests show that even the use of a microscopic GSF does not noticeably alter the NLD effects discussed in this work. We therefore adopted the default Kopecky–Uhl generalized Lorentzian model \citep{PhysRevC.41.1941} implemented in TALYS and kept it fixed in all NLD calculations. Finally, apart from the neutron-capture rates explicitly modified in this work, all other reaction rates were adopted from the default JINA Reaclib database \citep{cyburtJINAREACLIBDATABASE2010}.

\subsection{Sensitivity study} \label{subsec:sensitivity}

To quantify the impact of neutron-capture rates on rare-earth peak formation across different NLD models, we employed the $F$-metric, a standard approach in nucleosynthesis sensitivity studies. This metric evaluates the deviation in the final abundance distribution resulting from perturbations to individual reaction rates. For a specific nucleus $(Z, A)$, the sensitivity $F$ is defined as:
\begin{equation}
\begin{split}
F= 100 \sum_{A=150}^{180} \frac{|Y_{f}(A) - Y_{\text{base}}(A)| + |Y_{1/f}(A) - Y_{\text{base}}(A)|}{Y_{\text{base}}(A)},
\end{split}
\label{eq:sensitivity_measure}
\end{equation}
where $Y_{\text{base}}(A)$ is the baseline abundance using the original NLD model, and $Y_{f}$ and $Y_{1/f}$ denote the abundances after scaling the neutron-capture rate of nucleus $(Z, A)$ by a factor of $f$ and $1/f$, respectively. The summation covers the rare-earth mass region ($A=150-180$).

While previous sensitivity studies often employ large perturbation factors of one to two orders of magnitude \citep{surmanSensitivityStudiesWeak2014, mumpowerImpactIndividualNuclear2015}, \citet{mumpowerInfluenceNeutronCapture2012} demonstrated that a factor of merely 5 can largely alter the neutron capture channel for some nuclei, leading to flow saturation or exhaustion. To avoid drastic alterations to the baseline nuclear flow, which could smooth out the sensitivity differences between neighboring nuclides, we adopted a moderate factor of f = 3. This approach allows us to effectively investigate the sensitivity variations across different NLD models.

\section{Results and Discussion} \label{sec:results}
\subsection{Impact of NLD models on abundance distributions} \label{subsec:results1}

\begin{figure}[t!]
    \centering
    \includegraphics[width=1\linewidth]{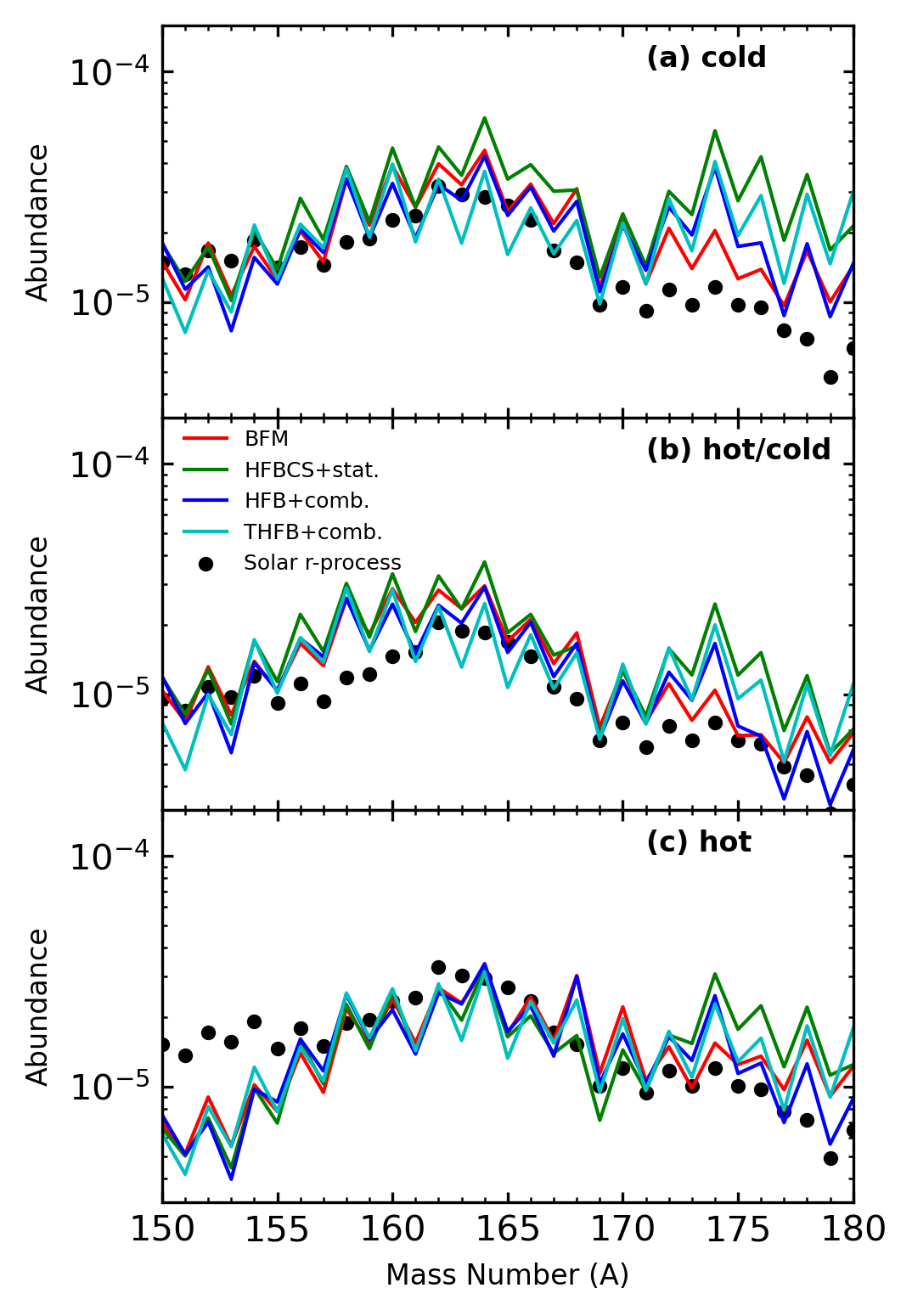}
    \caption{The final $\mathit{r}$-process rare-earth peak abundances simulated across three distinct astrophysical environments, calculated using neutron-capture rates derived from four different NLD models. Panels (a), (b), and (c) correspond to the astrophysical parameters: hot ($s=30, \tau=70, Y_e=0.18$), hot/cold ($s=20, \tau=10, Y_e=0.2$), and cold ($s=10, \tau=3, Y_e=0.2$). The lines represent the results using the BFM (red), HFBCS+stat. (green), HFB+comb. (blue), and THFB+comb. (cyan). The black circles show the solar $\mathit{r}$-process rare-earth peak abundance distribution, normalized to the average abundance of the $A=195$ peak calculated across the four models.}
    \label{fig:abundances}
\end{figure}

The final $\mathit{r}$-process abundance distributions in the rare-earth region ($A=150-180$) at $t = 10^{9}$ s, calculated using four different NLD models (BFM, HFBCS+stat., HFB+comb., THFB+comb.) across the three astrophysical scenarios, are displayed in Figure \ref{fig:abundances}. We use the BFM (the default NLD model in TALYS) to represent the phenomenological class and for the subsequent analysis, since we observe that the three phenomenological models produce remarkably similar abundance distributions, showing no strong structural deviations among themselves when compared to the significant variations observed among the three microscopic models. This consistency is expected: these phenomenological models are all based on the Fermi gas framework; they extrapolate the same experimental data using different fitted phenomenological formulae to describe the NLDs needed for the $\mathit{r}$-process.

\begin{figure*}[t!]
    \centering
    \includegraphics[width=1\linewidth]{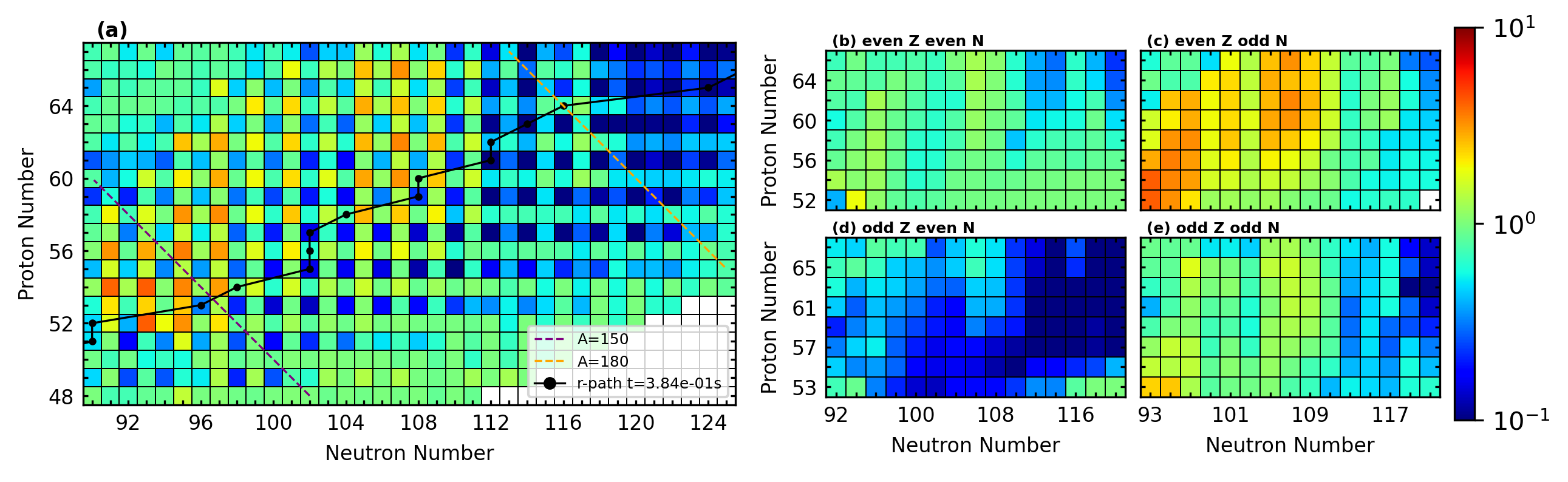}
    \caption{The ratio of the neutron-capture rates calculated using the HFBCS+stat. to those calculated using the BFM, covering the nuclear region $N=90-125$ and $Z=48-67$ at a temperature of $0.2 \text{ GK}$. Panel (a) shows the data for the complete range, with dashed lines marking the isobars $A=150$ and $A=180$. The labeled $\mathit{r}$-path in this figure traces the nucleus with the maximum isotopic abundance at $T=0.2 \text{ GK}$ (corresponding to evolution time $t=0.384 \text{ s}$), as determined by the nuclear network calculation utilizing the BFM in the cold scenario. Panels (b), (c), (d), and (e) focus on the central region ($N=92-121, Z=52-67$) and present the results broken down according to the even-odd parity combinations of the neutron and proton numbers.}
    \label{fig:rate_ratios}
\end{figure*}

We divide the analysis of the abundance distribution into two parts based on mass number. In the region $A=170-180$, the three microscopic NLD models produce a fake peak across all three scenarios. This is because these NLD models predict a global decline in NLD for nuclei approaching the neutron magic number $N=126$. This results in a lower neutron-capture rate in this region compared to the BFM (see Figure \ref{fig:rate_ratios}). When the $\mathit{r}$-process passes through this region, the nuclear flow encounters a bottleneck, resulting in the accumulation of material. This impact on the abundances varies depending on the dominant mechanism when the $\mathit{r}$-process path crosses this region. Due to the differences in the dominant nuclear physics inputs, the magnitude of the impact decreases progressively from the cold to the hot/cold, and finally to the hot scenario. Similar structures also exist in $\mathit{r}$-process simulations based on the SLy4 \citep{chabanatSkyrmeParametrizationSubnuclear1998} Skyrme energy density functional calculations reported by \citet{martinImpactNuclearMass2016}. This suggests that such systematic deviations near $N=126$ might be a common characteristic inherent to some current microscopic frameworks when predicting nuclear properties far from stability.

In the region $A=150-170$, the four models show a generally similar trend of abundance distributions within each astrophysical scenario. When compared to the solar $\mathit{r}$-process, these models fit best in the cold scenario. In contrast, there is an extended pile-up of abundances in the hot/cold scenario before the rare-earth peak at $A \sim 164$, resulting in a broader rare-earth peak. In the hot scenario, there is an abundance decline at $A \sim 150$, leaving the abundances in this region lower than the solar $\mathit{r}$-process abundances. Our results show that in this region, the general abundance distribution strongly relies on the astrophysical environment, while changes in the NLD models are insufficient to alter this property.

Regarding the specific impact of changing the NLD model in this region, we observe that the deviation is largest at the peak $A \sim 164$ in the cold and hot/cold scenarios. The structure of the abundance deviation is similar in these two scenarios. However, the impact is larger in the cold scenario than in the hot/cold scenario. This is because the $(n,\gamma) \rightleftharpoons (\gamma, n)$ equilibrium in the hot/cold scenario breaks down slightly later than in the cold scenario. Although neutron captures affect a similar region, the differences in neutron-capture rates have a similar but smaller effect. In the hot scenario, the abundance distributions are largely consistent across most of the mass region. However, a notable exception occurs near $A \sim 169$, where the HFBCS+stat. predicts a depletion not seen in the BFM results. This deviation is absent in the other two scenarios. This unique feature likely arises from the specific dynamics of freeze-out in the hot scenario. As the neutron density and temperature decline, the $(n,\gamma) \rightleftharpoons (\gamma, n)$ equilibrium is gradually broken. Since odd-$N$ nuclei typically possess lower $S_n$ than their adjacent even-$N$ neighbors, their equilibrium is disrupted earlier. Thus, the neutron-capture rates of odd-$N$ isotopes begin to influence the abundance evolution while the even-$N$ isotopes are still governed by equilibrium, leading to the observed ``odd-$N$ effect''. This mechanism, which creates a sensitivity pattern distinct from the other two scenarios, has been previously discussed in \citet{mumpowerInfluenceNeutronCapture2012}.

While the three microscopic models exhibit distinct features in the $A \approx 150-170$ range due to differences in effective nucleon-nucleon interactions (Skyrme versus Gogny forces) and pairing treatments, a detailed discussion of specific local variations is of limited value. Instead, our primary focus lies on the systematic structural divergence between microscopic and phenomenological approaches. We observe that the three microscopic models consistently predict higher NLDs for even-even nuclei and lower NLDs for odd-odd nuclei compared to the BFM. This clear structural difference has a significant impact on the evolution mechanism of the rare-earth peak. To illustrate this impact, we select the HFBCS+stat. as a representative case for comparison with the BFM in our subsequent analysis.

\begin{figure*}[t!]
    \centering
    \includegraphics[width=1\linewidth]{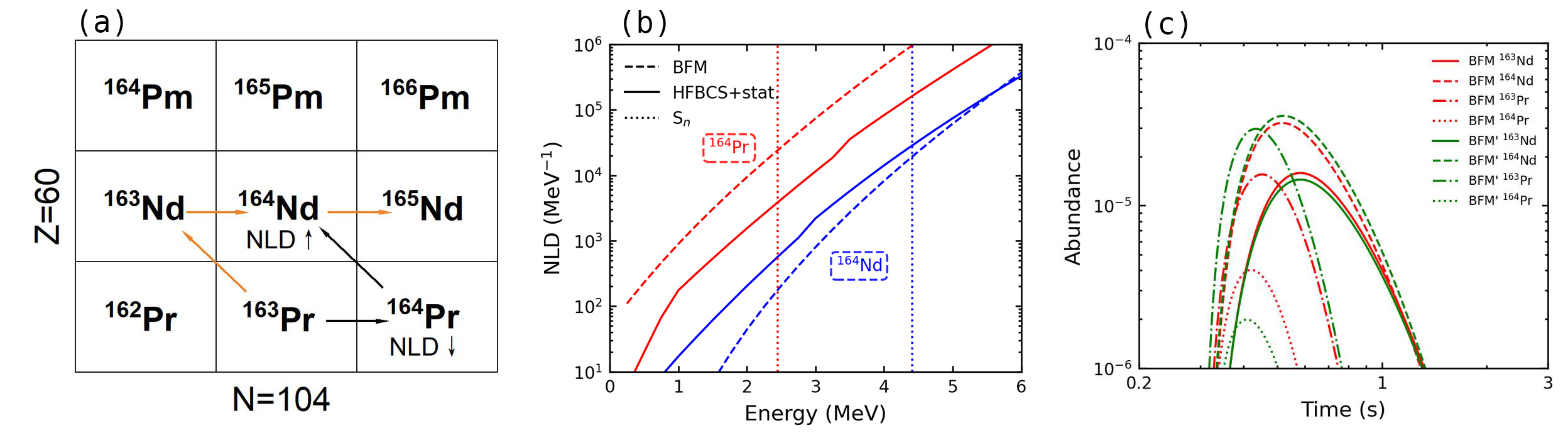}
    \caption{ Panel (a) illustrates the potential impact on the nuclear reaction network when the NLD of $^{164}\text{Nd}$ and $^{164}\text{Pr}$ are switched from the BFM to the HFBCS+stat. Arrows pointing up and left represent $\beta$-decay, while arrows pointing horizontally right represent neutron capture. Orange lines indicate that a reaction channel is relatively enhanced, and black lines indicate it is relatively suppressed. Panel (b) displays the NLDs predicted by the BFM and HFBCS+stat. for the neighboring even-even nucleus, $^{164}\text{Nd}$, and the odd-odd nucleus, $^{164}\text{Pr}$. The vertical lines mark the $S_n$ used in the FRDM mass model, which is adopted for calculating the corresponding neutron-capture rates. Panel (c) shows the differences in the time evolution of the abundances of $^{163}\text{Nd}$, $^{163}\text{Pr}$, $^{164}\text{Nd}$, and $^{164}\text{Pr}$ between $t=0.2$ and $3$~s in the cold scenario. These differences are obtained by comparing the original BFM model with the modified BFM$^{\prime}$ model, where the NLDs of $^{164}\text{Nd}$ and $^{164}\text{Pr}$ in the BFM have been specifically replaced with those from the HFBCS+stat.}
    \label{fig:toy_model}
\end{figure*}

The NLD does not enter the SkyNet reaction network directly; rather, it influences the network by altering the neutron-capture rates. To quantify the discrepancies between the two NLDs, we simply compare the resulting neutron-capture rates predicted by the HFBCS+stat. and BFM by defining the ratio:
\begin{equation}
\mathcal{R} = \frac{\langle \sigma v \rangle_{\text{HFBCS+stat.}}}{\langle \sigma v \rangle_{\text{BFM}}}.
\label{eq:rate_ratio}
\end{equation}

We found that in the cold scenario, the rare-earth peak forms at approximately 0.2 GK. Although neutron-capture rates are temperature-dependent, the relative discrepancy between the two models (quantified by the ratio $\mathcal{R}$) remains largely insensitive to temperature within the astrophysical window relevant to rare-earth peak formation. Therefore, we select a representative temperature of $T=0.2$~GK to analyze the rate ratios in the region $N=90-125$ and $Z=48-67$, as shown in Figure \ref{fig:rate_ratios}(a). The results reveal a pronounced sensitivity to the odd-even nature of nucleon numbers. To facilitate a detailed analysis, we decompose the central region of Figure \ref{fig:rate_ratios}(a) based on nucleon odd-even nature and display these subsets in panels (b)--(e).

Figure \ref{fig:rate_ratios}(a) demonstrates that beyond $N=112$, the capture rates based on HFBCS+stat. systematically exhibit a downward trend relative to the BFM predictions. This systematic suppression directly explains the abundance pile-up consistently observed in the $A=170-180$ region across all three scenarios in Figure \ref{fig:abundances}.

Figures \ref{fig:rate_ratios}(b)--(e) reveal that these discrepancies for even-$A$ nuclei in the $A=150-170$ region are significantly smaller than those for odd-$A$ nuclei. HFBCS+stat. predicts significantly higher reaction rates for even-odd nuclei compared to the BFM, whereas the trend is reversed for odd-even nuclei. The primary consequence of this systematic divergence is the redirection of nuclear flow and accumulation of even-even nuclei.

We present a simple case study around $^{164}\text{Nd}$ in Figure \ref{fig:toy_model}(a) to explain how NLD variations affect the abundance evolution. It is important to emphasize that the reaction rate differences in our study are driven directly by the NLD variations of the product nuclei $(Z, N+1)$. Therefore, we focus on the impact of even-$A$ NLDs, whose variations lead to significant deviations in the neutron-capture rates of the corresponding odd-$A$ target nuclei. Consider the nuclear flow starting from $^{163}\text{Pr}$. The HFBCS+stat. predicts a lower NLD below $S_n$ of the odd-odd nucleus $^{164}\text{Pr}$ than the BFM. This suppresses the reaction rate of $^{163}\text{Pr}(n,\gamma)^{164}\text{Pr}$. Since the $\beta$-decay rate remains unchanged, a portion of the nuclear flow is channeled to even-$Z$ isotopes through $\beta$-decay.

On the other hand, the HFBCS+stat. predicts a higher NLD for the even-even nucleus $^{164}\text{Nd}$. This enhancement increases the reaction rate of $^{163}\text{Nd}(n,\gamma)^{164}\text{Nd}$, facilitating the synthesis of $^{164}\text{Nd}$. However, since the destruction channels of $^{164}\text{Nd}$ remain relatively unchanged, it requires a higher abundance of $^{164}\text{Nd}$ to maintain the dynamical equilibrium during the abundance evolution. This local enhancement extends along the isotopic chain, further increasing the abundances across the entire chain.

Figure \ref{fig:toy_model}(b) compares the NLDs of the neighboring nuclei $^{164}\text{Nd}$ (even-even) and $^{164}\text{Pr}$ (odd-odd) as an example to describe the global NLD behavior of the even-$A$ nuclei.  Consistent with the observed trends in neutron-capture rates, the BFM predicts a higher NLD below $S_n$ for the odd-odd nucleus $^{164}\text{Pr}$ than the HFBCS+stat. However, for the even-even nucleus  $^{164}\text{Nd}$, the BFM predicts a lower NLD below $S_n$. The two models exhibit distinct behaviors due to their fundamentally different treatments of pairing correlations. The phenomenological BFM accounts for nucleon pairing via a simplified energy parameter $\Delta^{\mathrm{BFM}}$. It is defined as:

\begin{equation}
\Delta^{\mathrm{BFM}} = \chi \frac{12}{\sqrt{A}}+\delta,
\label{eq:BFM_pairing}
\end{equation}
where $\delta$ is a fitting parameter, and $\chi$ accounts for the odd-even effect ($\chi=0, 1, -1$ for odd-odd, odd-$A$ and even-even nuclei, respectively). This approach effectively treats the pairing energy as a constant rigid shift along the energy axis. The impact of this shift is highly pronounced in the low-energy behavior of the calculated NLDs (see the BFM predictions for $^{164}\text{Nd}$ and $^{164}\text{Pr}$ in Figure \ref{fig:toy_model}(b)).

The microscopic HFBCS+stat. provides a more physically reasonable description. It treats pairing correlations dynamically by solving the temperature-dependent BCS equations individually for each nucleus. In contrast, the treatment of pairing in the BFM has a much stronger impact on the level density, leading to a systematic and global discrepancy in the predicted NLDs of even-$A$ nuclei compared to the HFBCS+stat.

This systematic odd-even nature in NLD predictions directly reshapes the local nuclear flow. As demonstrated in Figure \ref{fig:toy_model}(c), we verify this mechanism using a hybrid test. We constructed a modified model, denoted as BFM$^{\prime}$, by replacing the NLDs of only $^{164}\text{Nd}$ and $^{164}\text{Pr}$ in the BFM with those from HFBCS+stat. (and updating the corresponding capture rates). We then calculated the abundance evolution for relevant nuclei in the cold scenario. Although changing only two nuclei has a limited impact, the local abundance evolution confirms our expectations: the reduced NLD of $^{164}\text{Pr}$ leads to an abundance increase of $^{163}\text{Pr}$ and an abundance decrease of $^{164}\text{Pr}$, while the enhanced NLD of $^{164}\text{Nd}$ directly increases its own abundance. 

\begin{figure}[t!]
    \centering
    \includegraphics[width=1\linewidth]{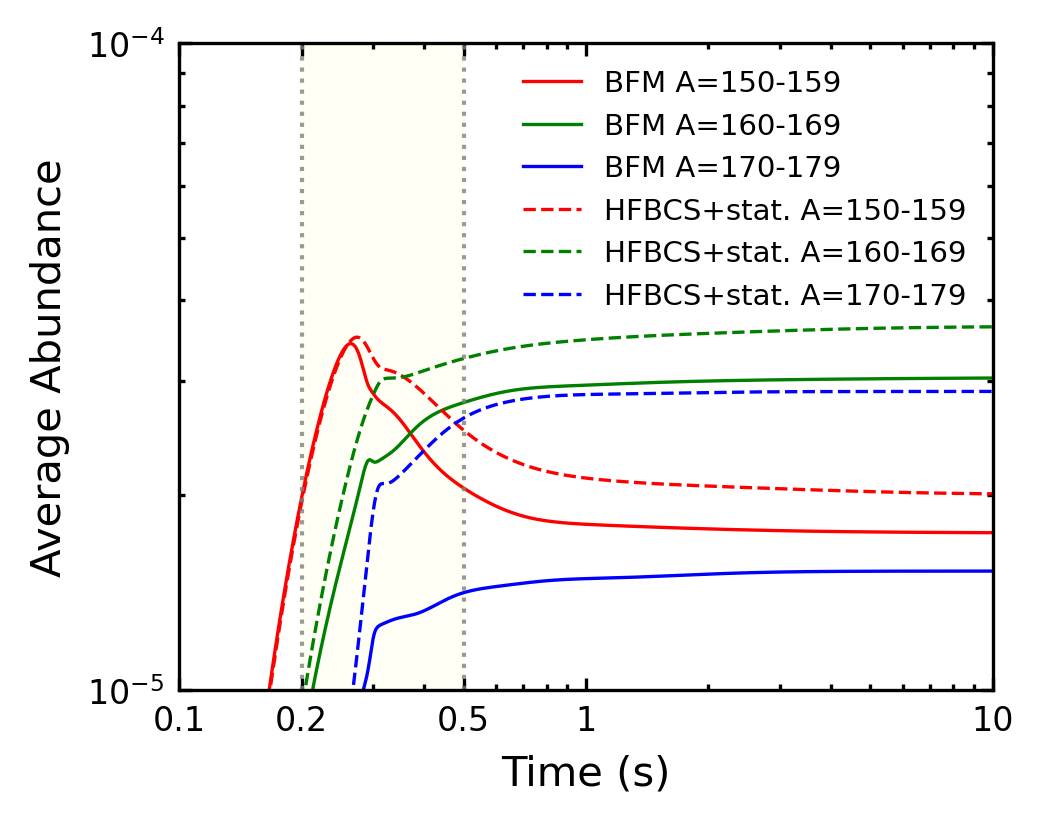}
    \caption{The time evolution of the average abundance in the three distinct regions of the rare-earth peak during the $\mathit{r}$-process in the cold scenario, as calculated using the HFBCS+stat. and BFM. The shaded light yellow area spanning from $t=0.2$ to $t=0.5$~s indicates the crucial time interval for the formation of the rare-earth peak.}
    \label{fig:rep_evolution}
\end{figure}

While Figure \ref{fig:toy_model} illustrates the microscopic impact of NLD variations on a typical even-even/odd-odd pair, Figure \ref{fig:rep_evolution} demonstrates the macroscopic result of these NLD changes on the entire rare-earth region. For this analysis, we divide the rare-earth peak into three mass regions: the left ($A=150-159$), central ($A=160-169$), and right ($A=170-179$) regions.

The transition from the phenomenological BFM to the microscopic HFBCS+stat. influences the rare-earth peak primarily through two mechanisms driven by changes in even-$A$ NLDs (and consequently, odd-$A$ capture rates). The first mechanism is the acceleration of rare-earth peak formation. The reduction in NLDs for odd-odd nuclei suppresses the neutron capture channel for odd-even nuclei, redirecting a larger fraction of the nuclear flow through the $\beta$-decay channel. This shift in the competition between capture and decay accelerates the upward progression of the nuclear flow along both $Z$ and $N$. This acceleration is evident in the central region of Figure \ref{fig:rep_evolution}, where the abundances predicted by HFBCS+stat. rise earlier and more rapidly than those of the BFM.

The second mechanism is enhanced material accumulation. Even-even nuclei are relatively stable and constitute the most abundant species in the reaction network. An increase in the NLD of even-even nuclei raises their production rate via neutron capture while their destruction channels remain almost unchanged. These results finally lead to material accumulation and a higher abundance along the $\mathit{r}$-process path. This effect strengthens the non-linear response of the system to neutron-capture rates. For instance, in the central region of Figure \ref{fig:rep_evolution}, HFBCS+stat. produces a noticeably higher abundance than the BFM. By contrast, in the left region, the increased proportion and stability of even-even nuclei slow down the decline in abundance.

Our results show that the differences in the odd-even effect between the BFM and HFBCS+stat. have a noticeable impact on the nuclear flow during the abundance evolution. While we focus on these two representative models for detailed discussion, we emphasize that such behaviors are also observed in other phenomenological and microscopic models within their respective classes. Since this systematic divergence is a global property inherent to the underlying frameworks, it is reasonable to expect a further impact on the sensitivity of the relevant nuclear physics inputs. Therefore, we extend our analysis to include all six NLD models in the subsequent NLD-based neutron-capture rate sensitivity study.

\subsection{Analysis of neutron capture sensitivity under different NLD models} \label{subsec:results2}

\begin{figure}[t!]
    \centering
    \includegraphics[width=1\linewidth]{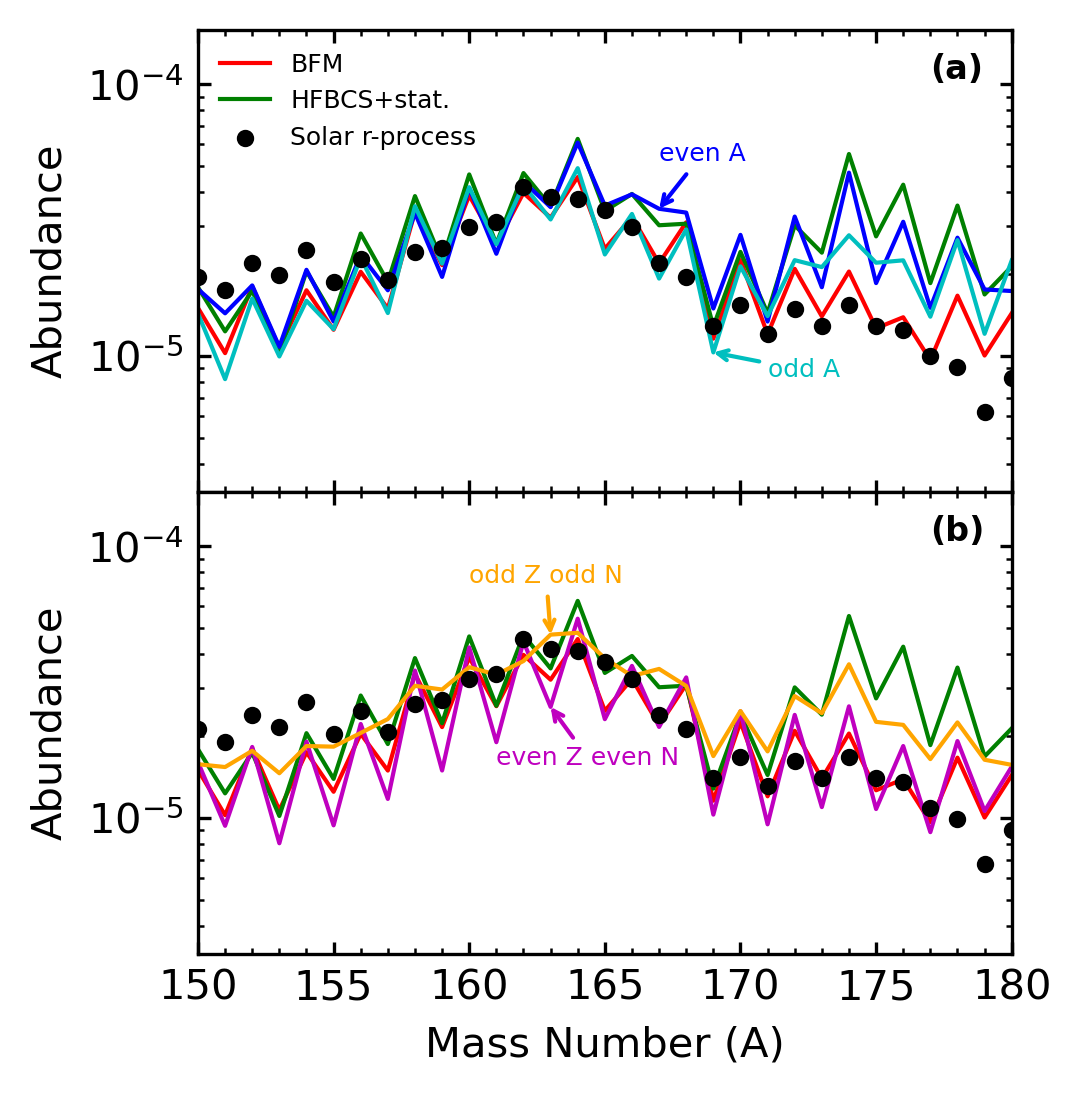}
    \caption{The final $\mathit{r}$-process rare-earth peak abundances calculated in the cold scenario. The results from the original BFM and HFBCS+stat. are included as references in both panels. Panel (a) shows the calculations where the NLDs of even-$A$ (even mass number) and odd-$A$ (odd mass number) nuclei in the BFM were substituted with the HFBCS+stat. NLDs. Panel (b) displays the results when the substitutions were specifically applied to even-$Z$ even-$N$ (even-even nuclei) and odd-$Z$ odd-$N$ (odd-odd nuclei) when computing the neutron-capture rates.}
    \label{fig:replacement_test}
\end{figure}

\begin{figure*}[t!]
    \centering
    \includegraphics[width=1\linewidth]{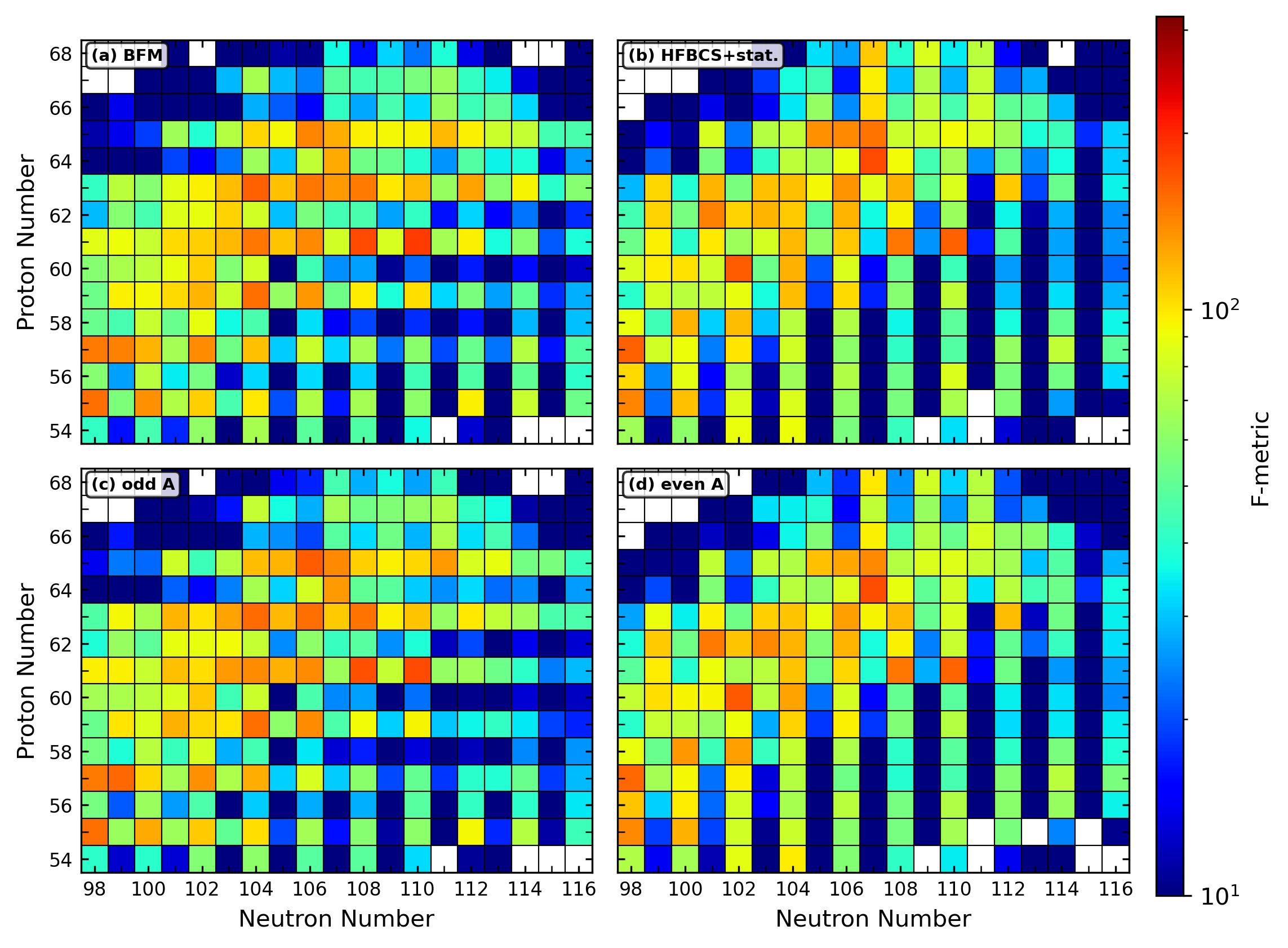}
    \caption{The sensitivity of the rare-earth peak abundances in the cold scenario to changes in the neutron-capture rate of a single nucleus. The color scale indicates the magnitude of the impact, increasing from blue (minimal) to red (maximal). White squares denote nuclei with negligible impact. Panels (a) and (b) show the results calculated using the neutron-capture rates derived from the BFM and the HFBCS+stat., respectively. Panels (c) and (d) show the results obtained by using the BFM as the baseline and replacing the NLDs of specific nuclear types with those from the HFBCS+stat. for the calculation of the neutron-capture rates.}
    \label{fig:sensitivity_measure}
\end{figure*}

Given that the discrepancies between the two models strongly depend on the proton and neutron odd-even nature, it is crucial to disentangle the specific contributions of different nuclear classes to the peak formation. To achieve this, we adopted a selective replacement strategy. Using the BFM as a baseline NLD model, we systematically replaced the NLDs of specific subgroups (e.g., all even-$A$ nuclei) with those from the HFBCS+stat., while keeping the others unchanged. We then recalculated the abundance distributions using neutron-capture rates derived from the corresponding NLD models and compared them with the full BFM and HFBCS+stat. results, as shown in Figure \ref{fig:replacement_test}.

The results in Figure \ref{fig:replacement_test}(a) reveal that replacing the NLDs of even-$A$ nuclei alone reproduces nearly the entire abundance difference observed between the full BFM and HFBCS+stat. In contrast, replacing only the odd-$A$ NLDs produces noticeable deviations only in the region $A=170-180$.

Given the pronounced impact of even-$A$ nuclei, we performed additional replacement tests by further separating them into even-even and odd-odd subgroups. We observed that the odd-even oscillation pattern of the abundance distribution is smoothed out in the odd-odd replacement results and enhanced in the even-even results. This occurs because the NLD differences of these nuclei are entirely one-sided, either consistently larger or smaller. Systematically changing the NLDs of only one specific group directly alters the reaction flow to adjacent nuclei of a different odd-even nature, inevitably disrupting the overall odd-even pattern of the entire rare-earth peak. Although the BFM and the HFBCS+stat. exhibit large discrepancies in even-$A$ NLDs, they both maintain consistent abundance oscillations. This is because the opposing impacts of the even-even and odd-odd NLDs mutually compensate for each other. These results demonstrate that a consistent and global treatment of nuclear structure is crucial for reproducing the correct shape of the rare-earth peak.

To preserve the internal consistency of the pairing correlation treatments during the selective replacement tests, we continue to employ the selective odd-$A$ and even-$A$ replacement strategy in the following sensitivity study. We computed the $F$-metric for the cold scenario by modifying the neutron-capture rates of nuclei in the mass range $A=140-190$ by a factor of 3, based on Eq. \ref{eq:sensitivity_measure}. This analysis was performed for the BFM, HFBCS+stat., and the corresponding selective replacement models. The results are presented in Figure \ref{fig:sensitivity_measure}.

\begin{figure*}[t!]
    \centering
    \includegraphics[width=1\linewidth]{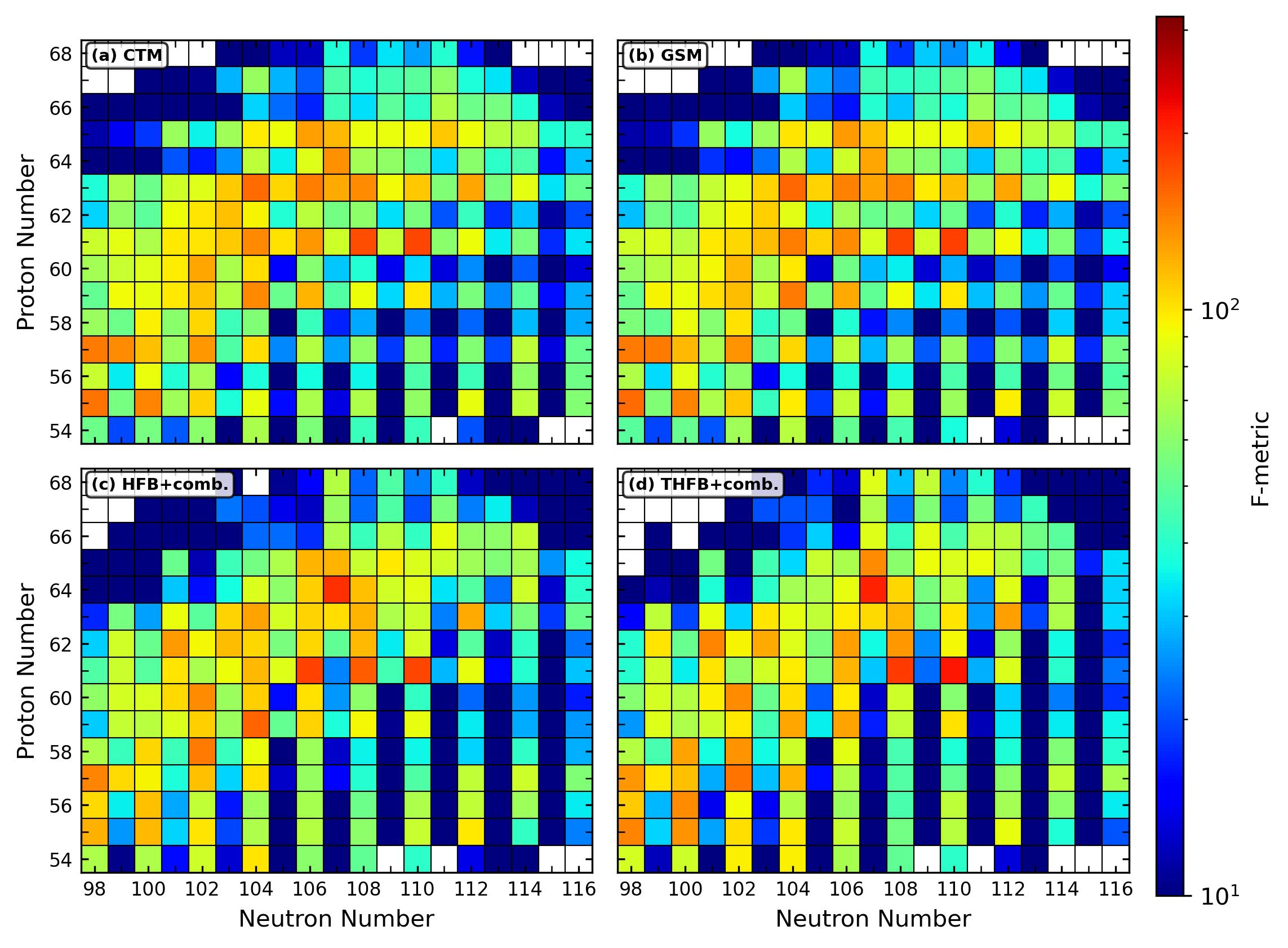}
    \caption{The sensitivity of the rare-earth peak abundances in the cold scenario to changes in the neutron-capture rate of a single nucleus, comparing four additional NLD models. The color scale used here is identical to that in Figure \ref{fig:sensitivity_measure}. Among these models, CTM and GSM are phenomenological models, while HFB+comb. and THFB+comb. are microscopic models.}
    \label{fig:other_models}
\end{figure*}

 The sensitivity map reveals that the relevant nuclei can be broadly characterized by two distinct physical regimes based on their distance from stability and the underlying nucleosynthesis dynamics:
\begin{itemize}
\item Regime I (roughly 15--30 neutrons from stability): In this highly neutron-rich region, the nuclear flow is primarily driven by a competition between neutron capture and $\beta$-decay. Consequently, the $F$-metric exhibits distinct features strongly correlated with nucleon odd-even nature.

\item Regime II (roughly 7--15 neutrons from stability): This regime corresponds to the progressive breakdown of equilibrium during the freeze-out phase. As the neutron abundance declines rapidly, the contribution of the neutron capture channel decreases, allowing $\beta$-decay to play a more prominent role. Under the combined effects of $\beta$-delayed neutron emission and neutron captures, the odd-even abundance oscillations are gradually smoothed out, thereby washing out the oscillatory pattern in the neutron-capture rate sensitivity within this regime. Meanwhile, given the proximity of this regime to stability, neutron captures have a much more direct impact on the final shape of the rare-earth peak, resulting in a higher sensitivity compared to Regime I.
\end{itemize}

It is worth noting that the boundary between Regime I and Regime II is not a rigid line, but rather an overlapping transition zone. As equilibrium breaks down gradually along the $\mathit{r}$-process path, the dependence of the $F$-metric on the odd-even nature observed in Regime I is progressively weakened as the flow enters Regime II. Furthermore, the exact locations and extents of these regimes are not absolute; they can be dynamically shifted by different astrophysical environmental conditions and variations in nuclear physics inputs.

\begin{figure*}[t!]
    \centering
    \includegraphics[width=1\linewidth]{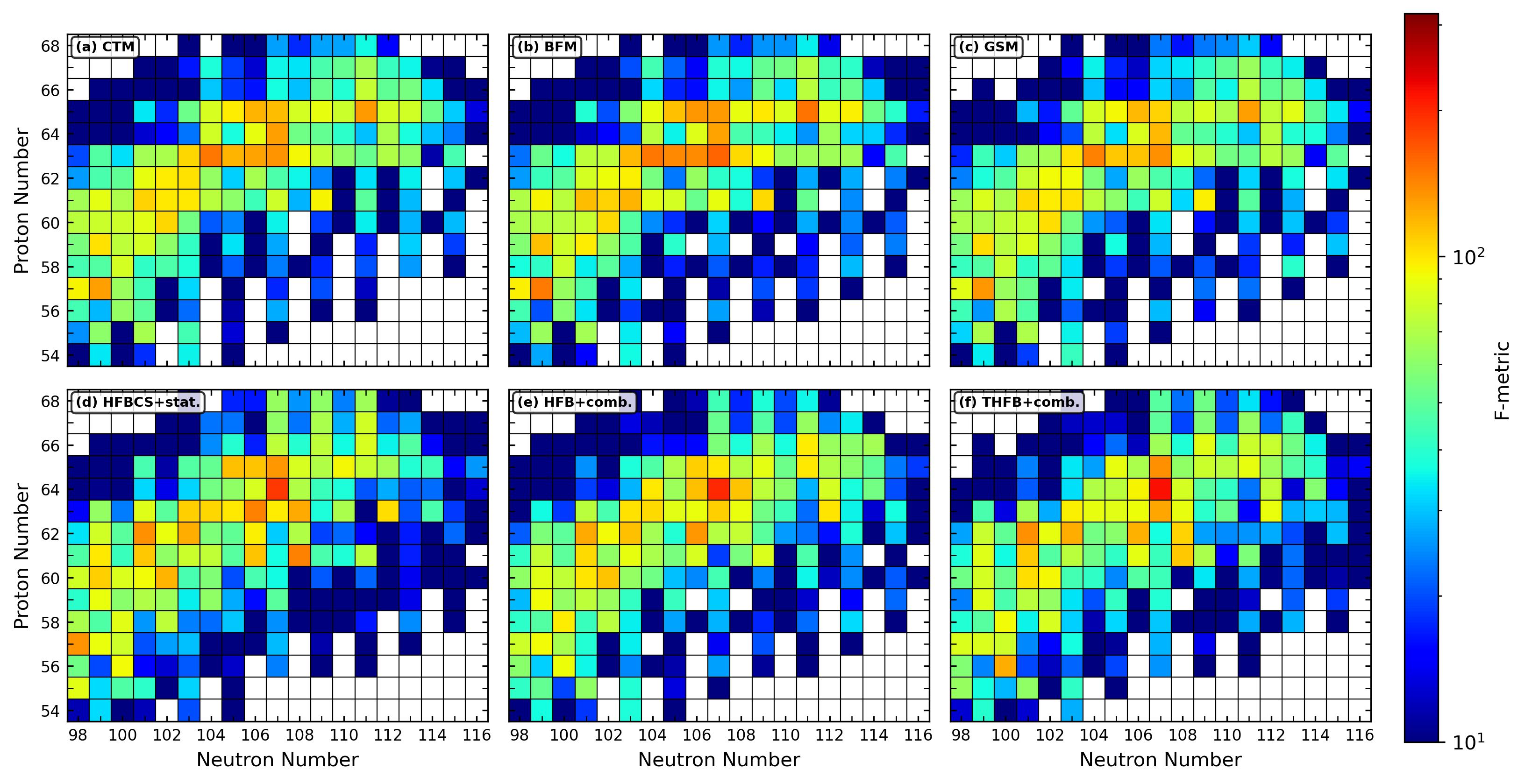}
    \caption{The sensitivity of the rare-earth peak abundances in the hot scenario to changes in the neutron-capture rate of a single nucleus, comparing the results derived from six different NLD models. The color scale used here is identical to that in Figure \ref{fig:sensitivity_measure}.}
    \label{fig:f_metric_hot}
\end{figure*}

Previous work by \citet{mumpowerInfluenceNeutronCapture2012} identified an ``even-$N$ effect'' in the cold scenario, where the neutron-capture rates of even-$N$ nuclei have a stronger impact than those of adjacent odd-$N$ nuclei. This effect is readily observed in Regime I of Figure \ref{fig:sensitivity_measure} across all models. While this even-$N$ dominance in the cold scenario contrasts with the ``odd-$N$ effect'' in the hot scenario, this previous study primarily focused on the sensitivity to the neutron odd-even nature and overlooked its correlation with the odd-even nature of protons. Indeed, our results reveal that the choice of NLD model further alters this sensitivity depending on the proton odd-even nature. As shown in Figure \ref{fig:sensitivity_measure}(a), for Regime I of the BFM results, the $F$-metric is highest for odd-even nuclei, followed by even-$A$ nuclei, and lowest for even-odd nuclei. In contrast, for Regime I of the HFBCS+stat. shown in Figure \ref{fig:sensitivity_measure}(b), even-$N$ nuclei possess significantly higher $F$-metric values than odd-$N$ nuclei. Meanwhile, the sensitivity to the proton odd-even nature is much less pronounced than in the BFM results.

The selective replacement analysis (Figure \ref{fig:sensitivity_measure}(d)) confirms that the variation in even-$A$ NLDs is the primary driver of these sensitivity shifts, while replacing odd-$A$ NLDs has a negligible effect (Figures \ref{fig:sensitivity_measure}(a) and (c)). This can be explained by the flow-redirect mechanism discussed in Figure \ref{fig:toy_model}(a). The transition from the BFM to the HFBCS+stat. involves an enhancement of even-even NLDs and a suppression of odd-odd NLDs. This structural change promotes the accumulation of even-even nuclei while depleting odd-odd nuclei. As a direct consequence, the $F$-metric of even-even nuclei in the HFBCS+stat. rises to a level comparable to that of adjacent even-odd nuclei, while the $F$-metric of odd-odd nuclei drops to match that of odd-even nuclei.  

Figure \ref{fig:other_models} shows the results for the remaining four NLD models. The phenomenological models and the microscopic models exhibit consistent sensitivity features within their respective groups. These results show that NLD variations between phenomenological and microscopic models lead to different neutron-capture sensitivity patterns. This demonstrates that NLDs contain distinct structural information essential for understanding the rare-earth peak formation.

Figure \ref{fig:f_metric_hot} presents the results of the six NLD models for the hot scenario. Although the abundance evolution mechanism differs significantly from the cold scenario, the affected mass area can still be divided into Regime I (roughly 15--30 neutrons from stability) and Regime II (roughly 7--15 neutrons from stability). In Regime I, all models exhibit a clear ``odd-$N$ effect''. In Regime II, due to a more rapid decline in neutron abundance and differences in early evolution mechanisms, the sensitivity is lower and exhibits a distinct pattern compared to the cold scenario, especially in phenomenological models. While we also investigated a hot/cold evolution scenario, the results are not shown here as they closely resemble those of the cold scenario.

\section{Summary and Outlook} \label{sec:summary}

In this work, we systematically investigated the impact of NLDs on the formation of the $\mathit{r}$-process rare-earth peak. By incorporating neutron-capture rates derived from six distinct NLD models into the SkyNet reaction network, we analyzed the abundance evolution across three representative astrophysical scenarios. Due to the high similarity within model classes, we selected the BFM and the HFBCS+stat. to represent the phenomenological and microscopic NLD models, respectively. We investigated the impact of their systematic differences from three perspectives: final abundances, abundance evolution mechanisms, and neutron-capture rate sensitivities. Our main findings are summarized as follows:

\begin{itemize}
    \item NLD impact on final abundances: The impact of discrepancies between NLD models on the final abundances can be divided into two distinct mass regions, with the magnitude varying depending on the astrophysical scenario. First, in the mass region $A=170-180$, all microscopic NLD models produce a fake peak across all three scenarios. This arises because microscopic models predict a global decline in NLD approaching the $N=126$ shell closure, creating a bottleneck that leads to material accumulation. Overall, the NLD impact gradually decreases in this region from the cold to the hot/cold, and finally to the hot scenario, a trend directly determined by when the $(n,\gamma) \rightleftharpoons (\gamma,n)$ equilibrium breaks down. Second, around the main peak at $A \sim 164$, noticeable abundance differences emerge in the cold and hot/cold scenarios. These variations are driven by systematic NLD discrepancies in even-$A$ nuclei between microscopic and phenomenological models. Conversely, in the hot scenario, the final abundances in this region remain almost unaffected across all models due to the persistence of the $(n,\gamma) \rightleftharpoons (\gamma,n)$ equilibrium.
    
    \item NLD impact on abundance evolution mechanisms: The systematic NLD deviations of even-$A$ nuclei between the HFBCS+stat. and the BFM exert a complex influence on the abundance evolution mechanism in the cold scenario. Compared to the BFM, the HFBCS+stat. globally predicts higher NLDs for even-even nuclei and lower NLDs for odd-odd nuclei. By suppressing the neutron-capture channels of odd-odd isotopes, this systematic deviation redirects the nuclear flow. Consequently, the flow enters the even-$Z$ isotopic chains earlier and accumulates on the more stable even-even nuclei. This underlying structural shift accelerates the early formation of the rare-earth peak and temporarily enhances its magnitude. These evolution behaviors are also observed in the remaining four phenomenological or microscopic NLD models.
    
    \item NLD impact on neutron-capture rate sensitivities: The NLD-induced alterations in the evolution mechanism fundamentally reshape the neutron-capture rate sensitivity map in the rare-earth peak formation zone (especially the region roughly 15 to 30 neutrons away from stability) in the cold scenario. In phenomenological models, the sensitivity exhibits strong oscillations that are highly dependent on the dual odd-even nature of both protons and neutrons. However, in microscopic models, their distinct even-$A$ NLD structure drives an abundance increase of even-even nuclei and an abundance decrease of odd-odd nuclei, resulting in a corresponding increase and decrease in their respective sensitivities. Ultimately, this effect completely eliminates the sensitivity variations associated with the proton odd-even nature in microscopic models. In contrast, the hot scenario does not exhibit such pronounced sensitivity differences between models, as its $\mathit{r}$-process path is primarily dominated by nuclear masses rather than neutron-capture rates.
\end{itemize}

In conclusion, our results demonstrate that the internal nuclear structure contained within NLDs plays a critical role in shaping the $\mathit{r}$-process rare-earth peak. Although NLD variations alter individual neutron-capture rates by a relatively small margin (mostly less than an order of magnitude), they collectively induce a global redirection of the nucleosynthesis pathway. Given that previous $\mathit{r}$-process studies have predominantly relied on phenomenological NLD models, the absence of explicit nuclear structure information likely introduces substantial uncertainties into current simulations. Therefore, future $\mathit{r}$-process simulations would greatly benefit from utilizing nuclear physics inputs derived from a unified microscopic framework. Specifically, evaluating NLDs, nuclear masses, and $\beta$-decay rates self-consistently using the same set of interaction parameters represents a crucial step toward reducing theoretical inconsistencies and better understanding the true origins of the $\mathit{r}$-process.

\begin{acknowledgments}
This work is also supported by Chinese Academy of Sciences Project for Young Scientists in Basic Research (YSBR-099) and the Key Laboratory of Nuclear Data Foundation (JCKY2025201C154).
\end{acknowledgments}

\bibliography{reference}{}
\bibliographystyle{aasjournalv7}

\end{document}